\documentclass[aps,prd,reprint,groupedaddress,longbibliography]{revtex4-1}

\usepackage{graphicx}	
\usepackage{array}
\usepackage{longtable}
\usepackage{amsmath}

\newcolumntype{L}[1]{>{\raggedright\let\newline\\\arraybackslash\hspace{0pt}}m{#1}}

\newcommand{\nuebar}{\bar{\nu}_e}
\newcommand{\antinu}{\bar{\nu}}

\begin{document}


\title{Sensitivity of seismically cued antineutrino detectors to nuclear explosions}


\author{Rachel Carr}
\email[Corresponding author:~]{recarr@mit.edu}
\affiliation{Laboratory for Nuclear Science, Massachusetts Institute of Technology, Cambridge, Massachusetts 02139, USA}

\author{Ferenc Dalnoki-Veress}
\affiliation{James Martin Center for Nonproliferation Studies, Middlebury Institute of International Studies at Monterey, Monterey, California 93940, USA}

\author{Adam Bernstein}
\affiliation{Nuclear and Chemical Sciences Division, Lawrence Livermore National Laboratory, Livermore, California 94550, USA}



\date{August 13, 2018}

\begin{abstract}
We evaluate the sensitivity of large, gadolinium-doped water detectors to antineutrinos released by nuclear fission explosions, using updated signal and background models and taking advantage of the capacity for seismic observations to provide an analysis trigger. Under certain realistic conditions, the antineutrino signature of a 250-kiloton pure fission explosion could be identified several hundred kilometers away in a detector about the size of the largest module currently proposed for a basic physics experiment. In principle, such an observation could provide rapid confirmation that the seismic signal coincided with a fission event, possibly useful for international monitoring of nuclear weapon tests. We discuss the limited potential for seismically-cued antineutrino observations to constrain fission yield, differentiate pure fission from fusion-enhanced weapon tests, indicate that the seismic evidence of an explosion had been intentionally masked, or verify the absence of explosive testing in a targeted area. We conclude that advances in seismic monitoring and neutrino physics have made the detection of explosion-derived antineutrinos more conceivable than previously asserted, but the size and cost of sufficiently sensitive detectors continue to limit applications.
\end{abstract}


\maketitle


\section{Motivation}

The possibility of detecting antineutrinos from a nuclear fission explosion was suggested at least as early as 1951, but such an observation has yet to occur. Plans in the 1950s \cite{Reines} and 1980s \cite{LANL1980s} to employ U.S. weapon tests in fundamental investigations of neutrinos were discarded in favor of experiments at nuclear reactors and accelerators. While recent work has explored antineutrinos as a nonintrusive tool for reactor surveillance (e.g., \cite{Christensen:2014pva}), the only publicly available assessment of antineutrino detection for explosion monitoring is nearly two decades old \cite{Bernstein2001}. That study evaluated prospects for detecting low-yield underground or underwater nuclear weapon tests missed by conventional methods of seismic, infrasound, hydroacoustic, radionuclide, and satellite-based monitoring. It concluded that the size and cost of antineutrino detectors able to identify these low-yield explosions exceeded practical limits at that time.

Large antineutrino detectors remain expensive and technically demanding, but two advances motivate an updated study. First, progress in basic neutrino physics has led to the proposal of detectors containing hundreds of kilotons of gadolinium-doped water in Japan and South Korea \cite{Abe:2011ts, Abe:2016ero}. The proposed detector sites are roughly 600 km and 900 km from the location of the six North Korean nuclear tests conducted since 2006. This regional proximity invites questions about whether the detectors could capture forensically useful signals in the event of a future North Korean test. Second, traditional explosion detection techniques have improved, especially through the construction of an international seismic network that now monitors most of the world with a very low detection threshold \cite{doi:10.1080/10736700.2016.1272207}. On one hand, this lowered threshold makes detecting subthreshold explosions with antineutrinos even more challenging than at the time of the previous study. On the other, the fact that the enhanced seismic network can now pinpoint suspected detonation times and locations opens new possibilities for antineutrino-based surveillance.

In this work, we consider antineutrino observation as a complement to established explosion monitoring techniques. Rather than targeting the very small yields examined in the previous study, we focus on explosions on the scale of 1 kton to hundreds of kilotons. We assume that a suspected fission event has been observed through established channels, most likely seismic sensing, and that the probable detonation time and location have been well constrained by these methods. The estimated detonation time, which we call a seismic cue, indicates when to look for an antineutrino signal in a detector datastream. This approach directs our attention to smaller event samples and more realistically sized detectors than those contemplated in the previous study.

We evaluate sensitivity for two cases of interest. Primarily, we consider the detector size and standoff distance required to confirm with high confidence that a seismic signal coincided with a fission event. This confirmation could play a role similar to the detection of radionuclides within the International Monitoring System overseen by the Preparatory Commission for the Comprehensive Nuclear-Test-Ban Treaty Organization; i.e., it could help to formally exclude alternative explanations of the seismic signal, such as earthquake or chemical explosion. Compared to radionuclides, antineutrinos have the advantages of appearing rapidly after an explosion and escaping from even fully contained underground explosions. As a secondary objective, we assess how well antineutrino observations could quantify the fission yield of an explosion. While the detection conditions would be very challenging to achieve, the comparison of yield estimates from seismic data and antineutrino observations could, in principle, help to discriminate pure fission from fusion-enhanced weapon tests, or to indicate that the explosive energy had been decoupled from the surrounding earth in an attempt to evade detection.

\section{Antineutrino production from a fission explosion \label{sec:prod}}

Electron antineutrinos ($\nuebar$) are emitted following fission of heavy nuclei such as $^{235}$U and $^{239}$Pu when the neutron-rich fission fragments undergo beta decay. In an average fission, each of the two fragments decays three times, leading to the emission of six antineutrinos. One kiloton of explosive yield (the equivalent of 1 kton of TNT) corresponds to approximately $1.45 \times 10^{23}$ fissions \cite{busterjangle} and thus the emission of approximately $8.7 \times 10^{23}$ antineutrinos from fission fragment decays. As weapon composition can vary and details are not publicly available, we base our calculations on a hypothetical highly enriched uranium, pure fission weapon in which all fissions occur on $^{235}$U. Similar calculations can be performed for other nuclei, such as $^{239}$Pu. In the case of $^{239}$Pu, studies of reactor antineutrino emissions suggest that the total antineutrino flux would be one third to one half that of $^{235}$U. The main fusion reactions that may be expected in boosted or thermonuclear weapons \cite{effects} do not produce antineutrinos or neutrinos. 

In addition to fission fragment decay, neutron capture and subsequent beta decays in fission fragments and surrounding material is an expected source of antineutrinos following a fission explosion. We do not model this contribution, since it depends on details of the weapon and environment. Including it in our model would increase the antineutrino flux expected for a given fission yield, but probably only modestly since the number of beta decays initiated in this way cannot be much larger than the number of fissions, and many of the emitted antineutrinos would be below detection threshold. Other features of fission explosions are less significant for the analysis presented here. Lorentz boosting of antineutrinos, even for fast-moving fragments in an explosion, has a negligible effect on the energy spectrum. Detections are likely to occur far enough from an explosion site that the underground explosion cavity can be approximated as a pointlike source.

Only approximate, one-dimensional estimates of the time and energy dependence of antineutrino emission from an explosion have so far been presented in the open literature \cite{Bernstein2001}. More attention has focused on the related case of antineutrinos produced in low-enriched uranium-fueled light-water-moderated fission reactors, which have been well measured in the context of neutrino oscillation measurements. We expect the energy spectrum of antineutrinos from a uranium- or plutonium-fueled fission explosion to be generally similar to reactor antineutrino emissions, with some differences due to the harder fissioning neutron spectrum in the explosion case. The time profile will be significantly different, with a rapidly decaying pulse in the case of an explosion versus the quasi-steady-state emission from a reactor.

For this study, we model the time and energy dependence of the antineutrino emission from a hypothetical uranium-fueled, pure fission explosion based on the following assumptions and approximations: all fissions occur on $^{235}$U within a microsecond of detonation \cite{pulseTime}; the fission-inducing neutron spectrum is unmoderated (the Watt spectrum \cite{Watt:1952zz}); all antineutrinos are produced by the beta decays of fission fragments and their daughters (neglecting neutron activation, as noted above); transitions from excited beta end states to the ground state are essentially instantaneous (the small population of isomers is neglected); and all beta decays have the simple allowed spectrum shape used in early reactor spectrum estimates \cite{vogel1981}. We neglect corrections to the beta spectrum shape since detailed energy spectrum information is not relevant for our sensitivity analyses.

Following these assumptions, we perform a Monte Carlo simulation of the beta decay chains of fission fragments. We begin with simulations of the instantaneous fission fragment yield from $^{235}$U by neutrons with kinetic energies of 2.0 MeV, the approximate mean of the Watt spectrum, supplied by the \textsc{freya} simulation package \cite{freya}. We simulate the beta decays of these fragments, and their daughters, until stability is reached, using data on half-lives, branching ratios, and endpoints extracted from the Evaluated Nuclear Structure Data File (ENSDF) database \cite{ensdf}. The ENSDF database is not entirely comprehensive, and we simply omit antineutrino emission from beta decays for which half-life or endpoint information is not available. Other nuclear databases could be substituted for ENSDF and the \textsc{freya} results, but given the approximations listed in the previous paragraph and the fact that we do not use detailed spectrum information in our ultimate calculations, the choice of database will have minimal impact on our conclusions.

\begin{figure}[ht!]
\centering
\includegraphics[width=1.1\columnwidth]{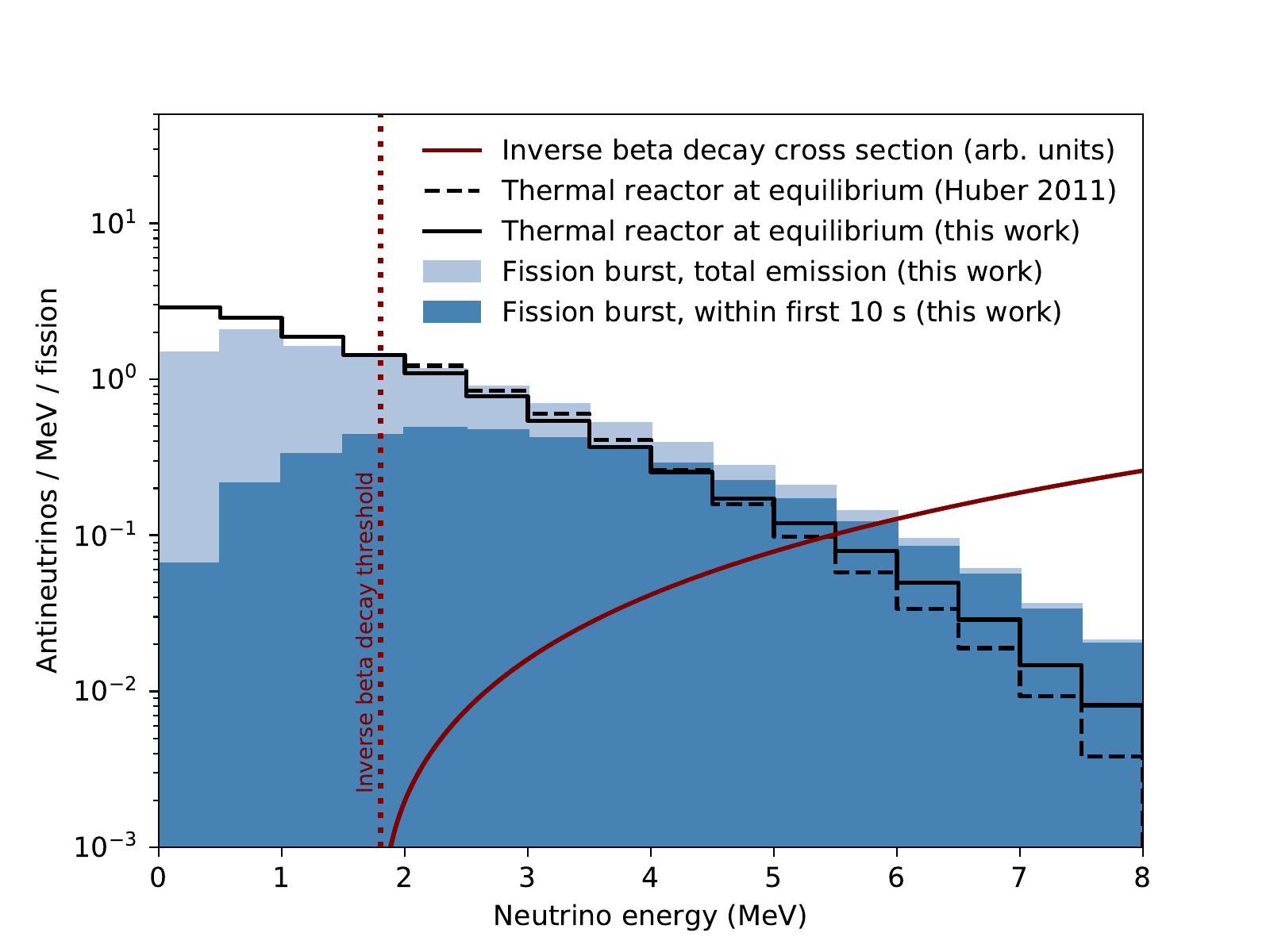}
\caption{The simulated energy spectrum of antineutrinos emitted from a $^{235}$U fission explosion, in total (light blue) and in only the first 10 seconds after detonation (dark blue). Overlaid are two estimates for antineutrino emissions from $^{235}$U fissions in a thermal reactor under equilibrium conditions: the estimate from our basic beta-branch summation model (solid black line) and one from a more sophisticated electron spectrum conversion model commonly used in neutrino oscillation studies \cite{Huber:2011wv} (dashed black line). The vertical dotted line marks the IBD threshold of 1.8 MeV, and the IBD cross section appears as a solid curve, shown here with arbitrary units.}
\label{fig:Enu}
\end{figure}

To benchmark our simulation, we use our model to estimate the energy spectrum from a thermal reactor at equilibrium. We compare this estimate to a more carefully modeled thermal reactor spectrum commonly used in neutrino oscillation studies \cite{Huber:2011wv}. Figure \ref{fig:Enu} includes both of these thermal spectrum estimates. Our estimated spectrum is somewhat harder than the reference spectrum, likely due to differences between our basic beta-branch summation method and the electron spectrum conversion method used for the reference. In the energy range relevant for detection (above the threshold marked in Fig. \ref{fig:Enu}, our spectrum includes about 10\% fewer antineutrinos than the reference spectrum, likely due to holes in the ENSDF database. We find these differences acceptable for the present study, since detailed spectrum shape information is not important for our study and underestimating the antineutrino flux would only make our sensitivity estimates more conservative.

\begin{figure}[t!]
\centering
\includegraphics[width=1.1\columnwidth]{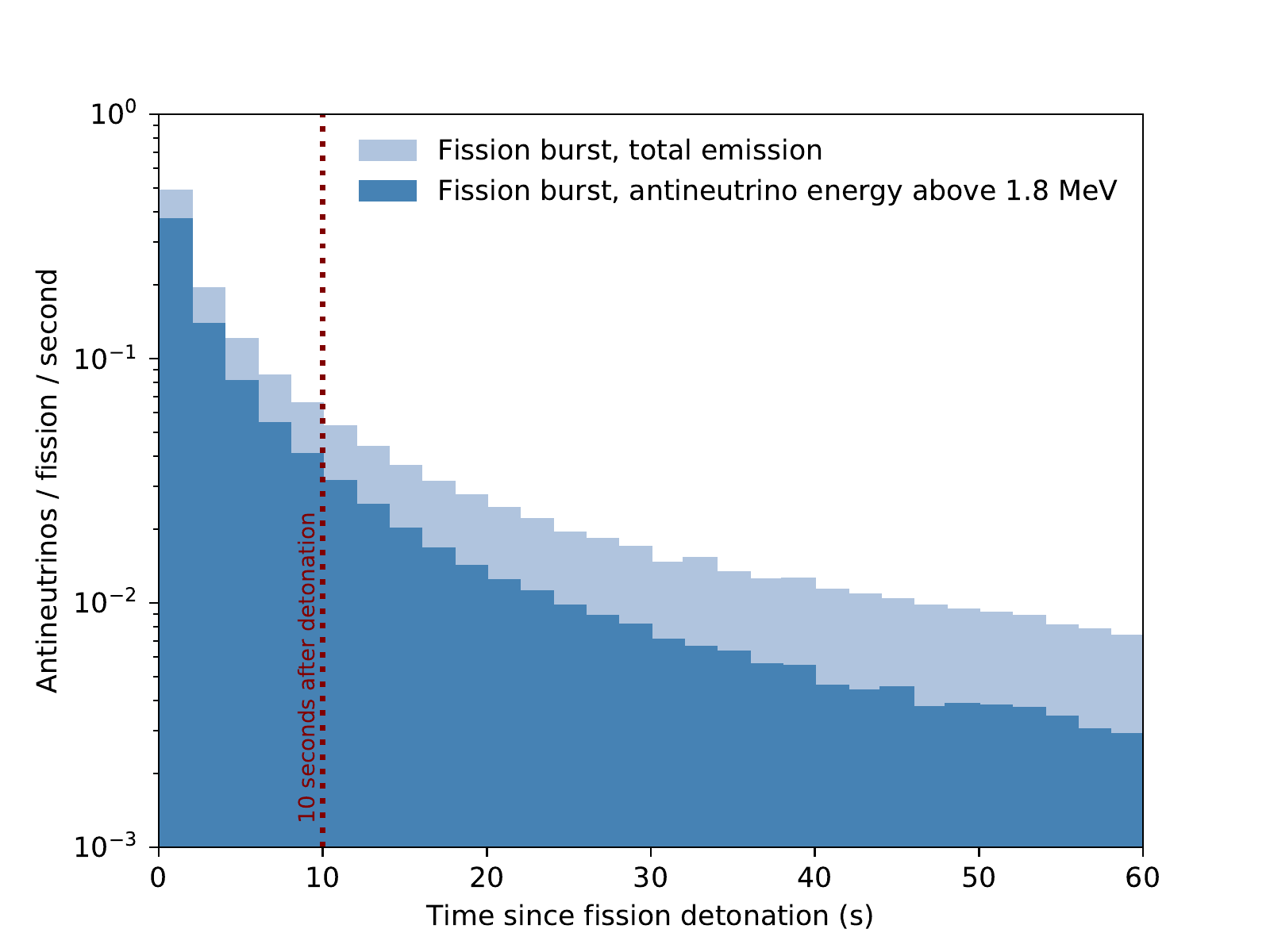}
\caption{The simulated time profile of antineutrinos emitted from a $^{235}$U fission explosion, normalized to one fission, for antineutrinos of all energies (light blue) and only for antineutrinos above the IBD threshold of 1.8 MeV (dark blue). The vertical dotted line marks 10 seconds post-detonation, the cutoff time for signals considered in this analysis.}
\label{fig:tAbs}
\end{figure}

The main result of our simulation is a two-dimensional model of antineutrino emission from a hypothetical $^{235}$U explosion, as a function of energy and time since detonation. A clear feature is the anticorrelation between emission time and energy expected for an ensemble of beta decays, with higher-energy antineutrinos tending to appear at shorter times. The one-dimensional projection in Fig. \ref{fig:tAbs} emphasizes the long tail of this distribution along the time axis. When all energies are considered, only about 30\% of antineutrinos are emitted within 10 seconds of detonation. However, most of the delayed emissions are below currently achievable detection thresholds. For antineutrinos with energies above 1.8 MeV (the energy threshold of the inverse beta decay detection channel discussed in Sec. \ref{sec:cue}), about 60\% of the flux is emitted within 10 seconds of detonation. 

Figure \ref{fig:Enu} shows the simulated energy spectrum of all antineutrinos emitted in the hypothetical explosion, along with the energy spectrum of only those antineutrinos emitted within the first 10 seconds following detonation. As noted above, this figure also includes our estimate of antineutrino emission from $^{235}$U in a thermal reactor at equilibrium. The total fission explosion spectrum is harder than the equilibrium thermal reactor spectrum. Further study would be needed to determined whether this difference is due primarily to the different energies of fission-inducing neutrons or the different abundances of beta decaying isotopes in a fission burst as compared to a reactor at equilibrium. More importantly for this analysis, the spectrum of antineutrinos emitted within the first 10 seconds of detonation has a considerably higher mean energy than that of the total emission, consistent with the energy-time anti-correlation noted above.

Based on the roughly 10\% normalization difference between our thermal spectrum model and the more precise reference, along with the difference in spectral shape, we estimate that systematic uncertainties on our simulated fission explosion spectrum are of the order of 10-20\% in the most relevant energy bins. This envelope is larger than the few-percent uncertainties on reactor antineutrino spectra calculated from reactor core simulations and nuclear databases. Our basic emission model could be improved through additional input from nuclear databases and more precise beta spectrum shape treatment similar to reactor analyses, but this level of precision suffices for the present discussion. 

\section{Observation in a seismically-cued water-based detector \label{sec:cue}}

Currently, the most feasible detection channel for antineutrinos from a fission explosion is the inverse beta decay (IBD) reaction, commonly used to detect antineutrinos from nuclear reactors: $\nuebar + p \rightarrow e^+ + n$. Coherent elastic neutrino-nucleus scattering has a higher cross section on neutron-rich targets, but as discussed in Sec. \ref{sec:coherent}, technology capable of detecting its signature for antineutrinos below 10 MeV is still under development.

Proton-rich detector media such as water and hydrocarbon-based liquid scintillator, instrumented with photomultiplier tubes or other light collectors, are ideal for IBD detection. While scintillator offers better detection efficiency and energy resolution, cost and environmental constraints make water a more practical choice for the large detector masses demanded in this application. Doping water with the gadolinium, which has been successfully demonstrated \cite{Dazeley:2008xk}, makes IBD events much more identifiable. These events create a double flash of prompt Cherenkov light from the positron track, followed within a few tens of microseconds by a gamma cascade arising from neutron capture on Gd.

Few other classes of events mimic the distinctive delayed coincidence signature of IBD interactions. Accidental coincidences of ambient radioactivity, cosmic-muon-generated fast neutron scatterings and captures, and cosmic-ray-produced beta-neutron emitters are the main non-antineutrino backgrounds. Their rates can be minimized by choosing radiopure detector materials and siting the detector deep underground. For weapon test monitoring, IBD interactions of reactor antineutrinos and geoneutrinos (antineutrinos produced by natural radioactivity in the earth) are also backgrounds. Their rates vary by geographic location, according to reactor proximity and local geology \cite{geoneutrinos}.

\begin{table*}
\centering
\renewcommand{\arraystretch}{1.2}
\begin{tabular}{
p{2cm}
p{1cm}
p{1.5cm}
p{2.5cm}
|
p{2cm}
|
p{1.5cm}
p{1.5cm}}
\multicolumn{4}{c|}{Detector} & \multicolumn{1}{c|}{Signal}& \multicolumn{2}{c}{Event counts in 10 s} \\
\hline
Fiducial mass (tons) & Depth (mwe) & Nearby reactors & Distance from explosion (km) & Fission yield (kton) & Signal $\nuebar$ & Total bkgd. \\
\hline
\hline
$1.9 \times 10^5$ & 2200 & Many & 600 & 250 & 2.3 & 0.7 \\
$5.0 \times 10^5$ & 2200 & Many & 900 & 250 & 2.5 & 1.3 \\
$5.0 \times 10^5$ & 2200 & Few & 200 & 10 & 2.5 & 1.3 \\
$1.0 \times 10^3$ & 270 & None & 10 & 10 & 2.9 & $<0.1$ \\
\hline
\end{tabular}

\caption{Estimated signal and background counts in a Gd-doped water Cherenkov detector, within a seismically cued 10 s window, for selected scenarios of detector mass, detector depth in meters water equivalent (mwe), prevalence of nuclear reactors in the region, distance from explosion to detector, and yield of fission explosion in kilotons.}
\label{tab:counts}
\end{table*}

One strategy for distinguishing a fission explosion signal from backgrounds is to require the signal to include many events in a short time window, such as ten events in 10 seconds, a grouping unlikely to occur by random coincidence of backgrounds. This strategy was the basis for the previous study of antineutrinos as a nuclear weapon monitor \cite{Bernstein2001}. The same technique is used to search for neutrino bursts from supernovae (which are in some sense a background for nuclear weapon test monitoring but not a real limitation, as sufficiently nearby supernovae occur rarely enough that the antineutrino flux is far below the flux from reactors and terrestrial radioactivity).

It is not necessary to require coincidence of many events if an external trigger, or time cue, can identify the window in which the signal should occur. If the window is short enough, the expected background count will be very close to zero and any observed events are likely to be signal, with a likelihood that can be easily quantified. For a suspected nuclear weapon test, the estimated detonation time derived from seismic observations provides an external trigger. We envision this trigger being applied at the analysis level to data that has been continuously recorded above a minimum signal threshold, as is typical for reactor neutrino detectors. We call this technique seismic cuing. The principle is similar to the common technique of requiring a beam trigger in accelerator-based experiments. It exploits the pulse-like time structure that inspired the original consideration of a fission explosion as an antineutrino source in the 1950s.

For suspected nuclear tests of a few kilotons or more, the detonation time can generally be inferred from seismic data to within approximately 2 s \cite{epicenter}. The length of the seismically cued signal window can be optimized for specific background levels, expected signal strength and standoff distance, and the available cuing precision. Optimization would balance the increased absolute signal rate in a longer window against the decreased signal-to-background ratio. Throughout this study, we use a 10 s window as a demonstration. This window is large enough to contain most of the observable antineutrino flux, according to Sec. \ref{sec:prod}, and significantly longer than the expected uncertainty on a seismic cue. According to simple optimization studies, 10 seconds is close to the optimal time window for the detection efficiency, background rates, yields, standoff distances, and detection criteria used in this study. While choosing a window shorter than 10 seconds would clearly increase the signal-to-background ratio, given the time structure of Fig. \ref{fig:tAbs}, the loss in absolute signal rate would result in an overall degradation of the sensitivity shown in Fig. \ref{fig:massDist}. Again, we note that a window length shorter or longer than 10 seconds may be optimal for efficiency, background, and standoff conditions different from the nominal cases we consider.

In a real application, the window would likely be opened a few seconds before the cue time to account for uncertainty in the seismic analysis. For simplicity, this study assumes that the cue occurs exactly at the time of detonation. The antineutrino transit time from source to detector, of the order of milliseconds or less for the distances we consider, is negligible in this analysis. Selecting the desired time window from a neutrino dataset is straightforward. Near-real-time analysis techniques developed for supernova triggers \cite{Antonioli:2004zb} suggest that data could be processed and analyzed almost as soon as a seismic time cue is available, likely before radionuclide analysis would be available to confirm the presence of fission.

\section{Estimation of observable signal}

Using the signal simulation described in Sec. \ref{sec:prod}, and following the detection scheme outlined in Sec. \ref{sec:cue}, the number $N$ of antineutrinos that could be detected from a fission explosion of yield $Y$ kton in a Gd-doped water detector of fiducial mass $M$ located a distance $L$ from the explosion site ($L$ is the chord connecting these points through the earth, not the great circle distance on the earth's surface, but these are negligibly different for distances of up to 1000 km) is as follows:
\begin{align}
N =~&\frac{p M N_A}{A}
\frac{fY\eta}{4\pi L^2} \nonumber \\
&\times \int_{E_{0}}^\infty dE 
\int_{t_0}^{t_0+\Delta t} dt \;
S(E, L) \phi(E, t)  \sigma (E)
\label{eq:signal}
\end{align}
In this expression, $p = 2$ is the number of free protons (hydrogen nuclei) per water molecule; $N_A$ is Avogadro's constant; $A$ is the molar mass of water; $f = 1.45 \times 10^{23}$ is the number of fissions per kiloton of explosive yield; $\eta$ is the detection efficiency for IBD events, discussed below; $E$ is the antineutrino energy; $E_{0} = 1.8$ MeV is the IBD threshold; $t$ is the time of antineutrino emission, integrated from the suspected detonation time $t_0$ for a signal window of length $\Delta t$; $S$ is the probability that an initially electron-flavored antineutrino of energy $E_\nu$ will be detectable in electron flavor after traveling the distance $L$, calculated using current global fits for neutrino flavor oscillation parameters \cite{Patrignani:2016xqp} (we neglect any possible oscillation to sterile neutrino states); $\phi$ is the simulated antineutrino flux; and $\sigma$ is the IBD cross section, for which we use a standard parametrization \cite{Vogel:1999zy, Pichlmaier:2010zz}. 

The detection efficiency parameter $\eta$ includes the ratio of detector livetime to total time in $\Delta t$, accounting for cosmic muon vetoes and any other deadtime, and the efficiency of all IBD selection cuts. For this study, we estimate that muon veto deadtime leads to an efficiency factor of 0.9, achievable with the detector depths and geometries that we envision. Deadtime from electronics and other factors is assumed to be negligible. The efficiency of large Gd-doped water detectors to IBD events below 10 MeV has not yet been experimentally studied, so we work from simulations performed by the WATCHMAN collaboration for a kiloton-scale detector with 0.1\% Gd loading and 40\% photocathode coverage, optimized for reactor antineutrino observation. These simulations indicate that signal detection efficiency of 65\% is achievable at the price of relatively high accidental-coincidence backgrounds \cite{Askins:2015bmb}. For our central value signal estimations, we use a more conservative 50\% signal detection efficiency over the range 1.8 MeV $< E_\nu <$ 10 MeV, approximated as energy-independent. Achieving this efficiency for volumes much larger than the WATCHMAN scale of 1 kton may require detector segmentation, increased photocathode coverage, or other enhancements. Table \ref{tab:counts} lists estimates for the number of observable antineutrinos in various scenarios of detector mass, depth, proximity to reactors, and distance from explosions of selected yields.

\section{Sensitivity for identifying fission explosions}
\label{sec:id}

Depending on background levels, observation of antineutrino interactions in a seismically cued signal window could provide a statistically credible answer to the question: Did the source of the suspect seismic signal involve nuclear fission? We quantify antineutrino-based sensitivity to this question with a simple hypothesis test based on counting statistics. The null hypothesis includes only background events. The alternative hypothesis includes both background and an antineutrino signal from a fission explosion. This test neglects systematic uncertainties on background rates, but we expect that these could be constrained to a level well below statistical uncertainties. We also leave out specific timing and energy information, as well as the weak, stochastic directionality information available for an ensemble of IBD events in water, as these would likely have minimal effect on sensitivity in small-signal scenarios. 

We perform this hypothesis test for a variety of explosive yields, detector-to-explosion distances, detector sizes, detector depths, and regional locations. To predict signal counts, we use Eq. \ref{eq:signal}. To predict counts from accidental, fast neutron, and cosmogenic isotope backgrounds, we scale estimates made by the WATCHMAN collaboration \cite{watchman}. To be conservative, we use background rates estimated for relatively loose selection cuts. We scale these rates by detector volume, or surface area in the case of radioactive contaminants in photomultiplier tubes, and we scale muon-induced backgrounds according to muon rate and energy variation with detector depth \cite{Mei:2005gm, muons}. We estimate the reactor antineutrino and geoneutrino background by choosing representative locations on a worldwide map of expected flux from both sources \cite{geoneutrinos}.

\begin{figure*}
\centering
\includegraphics[width=0.65\textwidth]{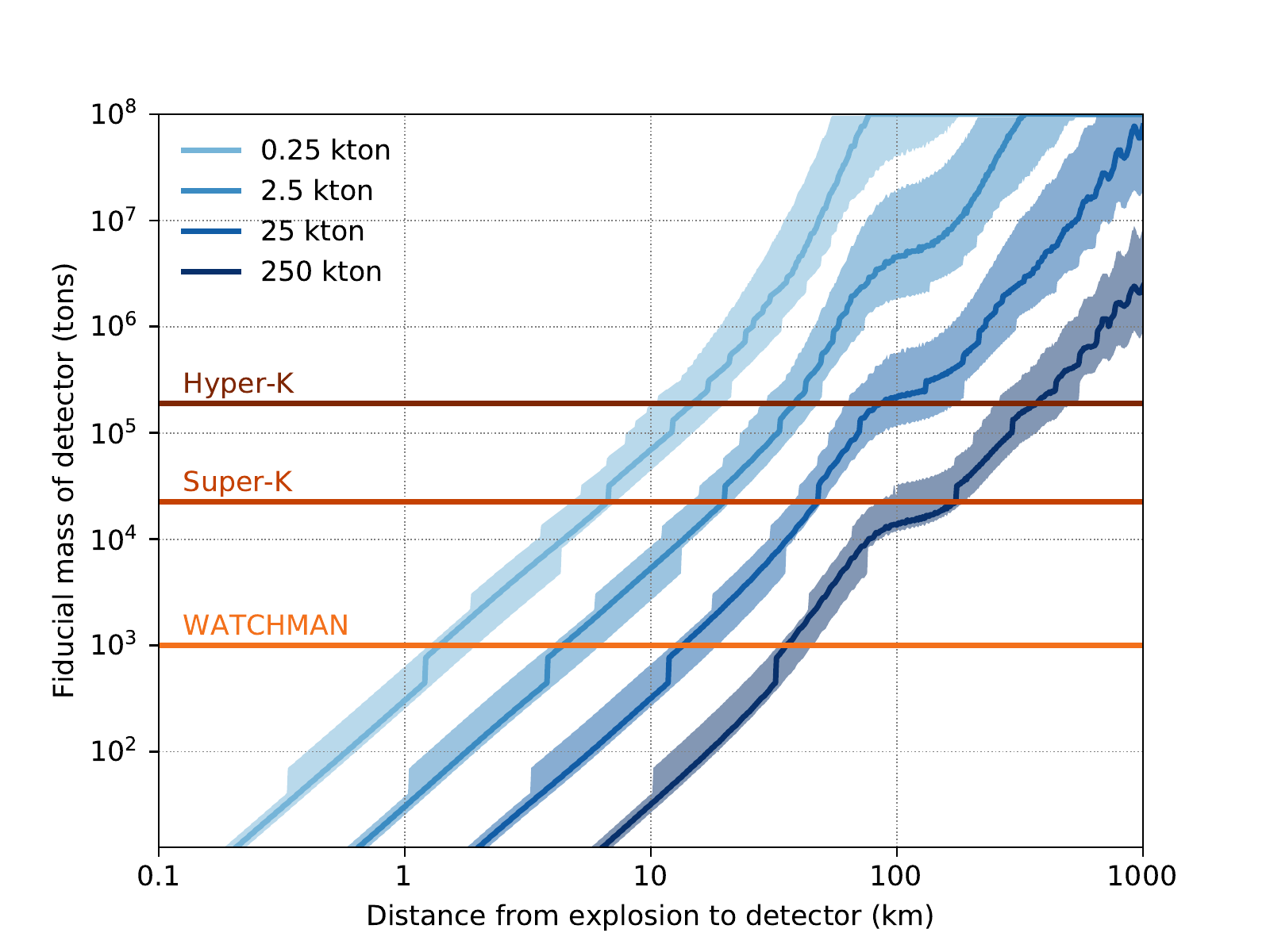}
\caption{The fiducial mass of a Gd-doped water detector required to achieve at least $90\%$ probability of positively identifying a fission event at $99\%$ or higher confidence level, as a function of detector distance from the explosion site, for varying fission yields in kilotons (kton). Central value estimates (solid blue curves) are based on 50\% signal detection efficiency and background rates scaled from WATCHMAN simulations, assuming a detector depth of 2200 meters water equivalent and many nearby reactors; these conditions match the first two rows in Tab. \ref{tab:counts}. Bands (shaded blue) cover signal detection efficiency ranging from 40\% to 60\% and background rates ranging from 0.2 times to 5 times the central value estimates. The standard fiducial mass of the largest existing neutrino detector (Super-Kamiokande, 22.5 kilotons), one tank of the largest proposed detector (Hyper-Kamiokande, 190 kilotons), and a moderately sized detector proposed for reactor monitoring (WATCHMAN, 1 kiloton) are superimposed (orange-brown lines).}
\label{fig:massDist}
\end{figure*}

Figure \ref{fig:massDist} is one way to represent sensitivity to positively identifying fission explosions. The figure indicates the size of a 2200-mwe underground, Gd-doped water detector that would be required to achieve at least $90\%$ probability of positively identifying a fission event at $99\%$ confidence level or greater, as a function of detector distance from the explosion site, for four different explosive yields. The central value curves assume 50\% signal detection efficiency and the background scaled from WATCHMAN as noted above. The bands cover scenarios ranging from 40\% to 60\% signal detection efficiency and backgrounds from 0.2 to 5 times the background rates scaled from WATCHMAN. The step discontinuities in these curves come from the small numbers of discrete signal events required in low-background scenarios. For example, the discontinuity just below the WATCHMAN line corresponds to the jump from needing one observed event to reach the desired confidence level to needing two events. Note that the scenarios of greatest interest involve detection of two or more events. The smooth waves come from flavor oscillations. Indicated in this figure are the fiducial masses typically quoted for the largest existing neutrino detector (Super Kamiokande, with a fiducial volume of 22.5 kilotons for most analyses), the largest proposed detector (one of the two tanks for Hyper-Kamiokande, with a proposed fiducial mass of 190 kilotons) \cite{HKdesign}, and a more moderately sized detector proposed for reactor monitoring (WATCHMAN, with a fiducial mass of 1 kiloton).

Under our central value model, in over $90\%$ of cases a detector the size of one proposed Hyper-Kamiokande tank (190 kilotons fiducial) would observe enough antineutrino events to positively identify, at greater than $90\%$ confidence level, a pure fission explosion of 250 kilotons occurring up to 450 km away. In over 80\% of cases, a detector of this size could identify a 250-kton explosion with at least 85\% confidence at a distance of up to 600 km. Estimates of the fission yield of North Korean nuclear test in September 2017 vary widely, with many around 250 kilotons \cite{norsar, 38north, popmech}, with no definitive statements about the fraction occurring from fission. The sites proposed for a Hyper-Kamiokande detector in South Korea are about 600 km from the North Korean nuclear test site, so if this detector had already been built and was operating, it may have had an opportunity to rapidly confirm that the September 2017 seismic event involved nuclear fission. Detectors on the scale of the largest existing detector, Super-Kamiokande, would have a high probability of confirming fission explosions at closer range, including distances up to slightly over 100 kilometers for large yields.

Detectors of a more moderate size, similar to the scale of WATCHMAN, would have a high probability of positively identifying fission explosions down to about 10 kton of yield within about 10 km. An application for this capability could be deployment of kiloton-scale detectors in specific sensitive areas to demonstrate that fission explosions were not occurring there, at least above the relatively large yield threshold of 10 kton. For example, as part of a cooperative monitoring agreement, a nation could construct a detector on a former nuclear test site. Seismic activity in that area could be checked against the antineutrino data stream, possibly reducing the need for on-site inspections in an area which may still be active for reasons other than nuclear testing. 

For all of these cases, we note that the detector size required for meaningful sensitivity is smaller than that suggested in previous, low-yield-focused work \cite{Bernstein2001} but probably still too large to be widely deployable. As a rough cost scale, the Super-Kamiokande detector required about \$100 million to construct in the 1990s \cite{nytimes}. The cost per ton for a Gd-doped detector with high photocathode coverage, built today, would likely be higher.

\section{Sensitivity to fission yield \label{sec:yield}}

\begin{table*}[ht!]
\centering
\renewcommand{\arraystretch}{1.2}
\begin{tabular}{
L{3.5cm}
L{3.5cm}
L{5.5cm}
}
Distance from explosion to detector (km) & True yield of fission explosion (kton) & Most probable 68\% CL interval for $\antinu$-based yield measurement (kton) \\
\hline
600 & 250 & 170--330 \\
200 & 50 & 40--60 \\
\end{tabular}
\caption{The most probable 68\% confidence level (CL) intervals on an antineutrino-derived measurement of explosive yield from fission for various scenarios of detector fiducial mass, distance from an explosion, and true fission yield in kilotons, based on statistical uncertainty only. The detector is a 1 megaton fiducial, Gd-doped water Cherenkov detector with moderate backgrounds, as estimated for a detector located at a depth of 2200 meters water equivalent in a region with many reactors.}
\label{tab:yield}
\end{table*} 

Since the number of antineutrinos emitted in a fission explosion is directly proportional to the number of fissions, a measurement of antineutrino flux provides a constraint on the explosive yield from fission. More specifically, the constraint is on the factor $S(L)Y/L^2$, but for simplicity we assume that $L$ is perfectly known from seismic data. Depending on proximity to seismic sensors, explosion epicenters generally can be well inferred \cite{epicenter}. We quantify sensitivity for yield measurements based on simple counting statistics, as in the previous section. Again, we neglect systematic uncertainties on the signal model, since statistical uncertainty dominates and systematics could likely be reduced by more rigorous modeling. Tightly constraining yield estimates requires more antineutrino events than merely confirming that some fission occurred, making detector masses requirements for strong yield constraints larger than those identified in the previous section.

Table \ref{tab:yield} shows the most probable 68\% confidence level interval on a fission yield measurement for two example scenarios: a relatively high-yield explosion observed from a relatively long distance and a smaller yield observed from a shorter distance, both in a megaton-scale detector (where most probable interval means the interval derived in the case where the mean expected number of events is observed). These intervals are of a similar magnitude to uncertainties reported on some seismic yield estimates, which are limited by uncertainties about the test site geology and configuration, particularly uncertainty on the underground depth of the explosion \cite{dahlman}.

Antineutrino-based yield measurements may be most useful in cases where they disagree significantly with yield estimates from seismic data or other observations based on ground movement, such as radio interferometry. An antineutrino-based yield estimate that is significantly larger than the apparent seismic yield could indicate that the test site cavity had been engineered to reduce explosive energy coupling to the surrounding earth. Estimates suggest that well-engineered cavity decoupling could reduce the apparent yield of a test, as inferred from seismic data, by a factor of up to 70 for yields up to a kiloton and 10-20 for some higher yields \cite{NAP12849}. As Tab. \ref{tab:yield} indicates, a suitably sized and positioned antineutrino detector may be capable of measuring yield in this range with an error much smaller than a factor of 10. In such cases, as long as seismic yield estimates are reasonably precise, an antineutrino measurement could be an indicator of decoupling.

In principle, an antineutrino measurement that is significantly lower than a well-constrained seismic yield estimate could indicate that some fraction of the explosive yield came from fusion rather than fission. As noted in Sec. \ref{sec:prod}, the main fusion reactions expected in weapons do not produce neutrinos or antineutrinos. A deficit in antineutrinos, compared to the expectation from a seismic yield estimate, could therefore be evidence that the source event was not a pure fission event. It would be very difficult to discriminate between pure fission and fusion-boosted fission devices, in which the fusion reactions serve primarily to provide fission-inducing neutrons rather than to directly increase yield \cite{effects}. Differentiating pure or boosted fission devices from thermonuclear devices would be more feasible, as the latter may obtain up to about half their explosive yield directly from fusion \cite{effects}.

The above discussion is, of course, somewhat idealized. We give separate discussions of seismic decoupling (which reduces the seismic signal, relative to the antineutrino signal) and fusion-enhanced devices (which increase the seismic signal, relative to the antineutrino signal), but these effects could occur together in the same scenario, adding some ambiguity to the joint interpretation of seismic and antineutrino data, especially in low-signal situations. Additional ambiguity comes from the fact that different fission fuel compositions produce antineutrino rates varying by up to about 40\% \cite{vogel1981}. Thus, the analyses we discuss would be most definitive in cases where certain other facts are known. For example, a country may be known to have no sites suitable for seismically decoupled tests, or known to possess capabilities for uranium enrichment but not plutonium production. In any case, the antineutrino signal carries information distinct from the seismic data and can therefore provide further insight, as long as the seismic and antineutrino signals are measured with sufficient precision.

\section{Alternative antineutrino detection technologies \label{sec:coherent}}

We focus on IBD detection in monolithic Gd-doped water Cherenkov detectors as the most technically mature option for observing antineutrinos from fission explosions. Some potential enhancements and alternative detection channels merit brief discussion. The detection efficiency of a Gd-doped water detector could be increased with use of a water-based liquid scintillator and more advanced light collection systems, both under development \cite{Bignell:2015oqa, Adams:2016tfm}. At most, these enhancements could improve fission confirmation and yield sensitivity by about a factor of two, since the main limitation in our present analyses is raw number of signal events rather than detection efficiency.

Coherent elastic (anti)neutrino-nucleus scattering is an alternative detection channel which could increase the number of raw signal events substantially, compared to IBD detection. This process was recently observed for the first time \cite{coherent} with neutrinos of higher energy than those from a fission explosion and with a more clearly exploitable time signature. Technology capable of detecting fission-produced antineutrinos is under development (e.g., \cite{Billard:2016giu}). Coherent scattering detectors would observe more signal events per unit mass than IBD detectors, but this potential should be balanced against the expectation that viable detector media, such as cryogenic noble elements and bolometric crystals, are less scalable than water.

\section{Conclusions}

We present a simple model for the time- and energy-dependent emission of antineutrinos from a hypothetical $^{235}$U fission weapon. Using this model, and using reasonable assumptions about detection efficiency and backgrounds, we estimate the size of Gd-doped water detectors required to confirm that a suspect seismic signal coincided with a fission event. We have also explore sensitivity for antineutrino-based fission yield measurements, which in some extreme cases could identify seismically decoupled explosions or distinguish pure fission weapons from weapons with a significant fraction of energy from fusion. In general, the main limit on sensitivity is the raw number of signal events, with detection efficiency and background rates being less critical limitations.

Overall, we show that antineutrino-based nuclear weapon test monitoring has broader potential than previously suggested but remains at the edge of conceivable detection capabilities. In our view, it remains challenging to envision building very large-scale detectors specifically for this purpose. However, nuclear test monitoring or test ban verification could be considerations in the siting, design, or operation of detectors built primarily for basic physics, particularly as these detectors grow in size and capabilities.

\section{Acknowledgements}

We gratefully acknowledge the assistance of Oluwatomi Akindele, Marc Bergevin, Amelia Trainer, and Ramona Vogt, and the support of the MIT Pappalardo Fellowship in Physics. Lawrence Livermore National Laboratory is operated by Lawrence Livermore National Security, LLC, for the U.S. Department of Energy, National Nuclear Security Administration under Contract DE-AC52-07NA27344. The views expressed herein are not necessarily those of the NNSA or the U.S. government.

\bibliography{AntinuExp}

\end{document}